\def\year{2020}\relax
\documentclass[letterpaper]{article} 
\usepackage{aaai20}  
\usepackage{times}  
\usepackage{helvet} 
\usepackage{courier}  
\usepackage[hyphens]{url}  
\usepackage{graphicx} 
\usepackage{graphicx}  
\title{Behavioral Modeling of Persian Instagram Users to detect Bots}
\author{Muhammad Bazm , \Large \textbf{Masoud Asadpour}\\ 
Faculty of Electrical and Computer Engineering, University of Tehran\\ 
Northern Kargar st, Enghelab Square, Tehran, Iran\\
m.bazm@ut.ac.ir 
}
 
\begin{document}

\maketitle

\begin{abstract}
Bots are user accounts in social media which are controlled by computer programs. Similar to many other thing, they are used for both good and evil purposes. One nefarious use-case for them is to spread misinformation or biased data in the networks. There are many pieces of research being performed based on social media data and their results’ validity is extremely threatened by the harmful data bots spread. Consequently, effective methods and tools are required for detecting bots and then removing misleading data spread by the bots.
In the present research, a method for detecting Instagram bots is proposed. There is no data set including samples of Instagram bots and genuine accounts, thus the current research has begun with gathering such a data set with respect to generality concerns such that it includes 1,000 data points in each group.
The main approach is supervised machine learning and classic models are preferred compared to deep neural networks. The final model is evaluated using multiple methods starting with 10-fold cross-validation. After that, confidence in classification studies and is followed by feature importance analysis and feature behavior against the target probability computed by the model. In the end, an experiment is designed to measure the model’s effectiveness in an operational environment. Finally, It is strongly concluded that the model performs very well in all evaluation experiments.
\end{abstract}

\section{Introduction}
\noindent Before web usages become widespread, there were rather simple methods for communication among people, including letter, telephone, and fax. After this time, tools such as email and video chat platforms, such as Skype, appeared. These platforms were only capable of a peer to peer data transmission which was also limited by acceptable data types at the beginning of the era.

This generation was obsoleted by \textbf{Social Networks}; a new platform had brought novel facilities for \textbf{information sharing} and \textbf{connectivity}. The data shared by users were preserved and the relationship's networks were explicitly defined and accessible, thus a large amount of data, representing people's activities, was stored. This data is commonly referred to as \textbf{The Big Data}.

With the big data, new opportunities have appeared for solving problems with no exact solutions. Most of these problems were coming from social science creating a new field of study known as \textbf{Computational Social Science}. Multiple issues are threatening the validity of research results in this field among which \textbf{data pollution} is a major threat. One of the resources for this problem is an entity known as \textbf{Bot}\footnote{Computer Robot}. 

Bots are user accounts controlled by a computer program aimed at gathering or spreading data. Although they might be used for good purposes, a large number of them are used for evil goals. Accordingly, providing tools for detecting such accounts and canceling their effects is extremely required. In this research, an effective solution is suggested for this problem which has achieved great scores from multiple evaluation metrics. 

Here, Instagram is the target social media. Recently, it has become common among people from all around the world. Persian speaking users have a large community in this media such that analyzing their data provides valuable answers to important questions. 

\section{Literature Review}
\noindent Bot detection is a well-known problem in computational social science. In this section, the most important researches in this field are reviewed. They are divided into three categories, corresponding to the approach they have adopted. Each category is described below:

The first category includes: \cite{6280553},
\cite{Gao2015SybilFrameAD}, \cite{DBLP:journals/tifs/GongFM14}, \cite{Carminati2013}, \cite{Danezis_sybilinfer:detecting}, \cite{4542826}, \cite{4531141}, \cite{4767341}, \cite{DBLP:journals/corr/abs-1301-6725}, \cite{10.1145/1879141.1879191}, \cite{Leskovec08communitystructure}, \cite{10.1145/1851275.1851226}, \cite{7918016} and \cite{8023129}, \cite{6fea7744517d4694ad9dc535b9f37dc7}. In mentioned researches, the main approach is using network structures; as a result, their job includes mostly graph-based methods. The major weakness of this researches is their poor results compared to the last category.

In the second category: researches such as \cite{10.1145/2517040}, \cite{10.1145/2556609}, \cite{Wang2013SocialTT}, \cite{7490315} and \cite{10.1145/3110025.3110091} are found. Their research mostly relies on crowd-sourcing. Threatening user's privacy is the most important challenge in researches in this section.

In the last category researchers such as: \cite{6280553}, \cite{7490315}, \cite{10.1145/2556609}, \cite{Lee11sevenmonths}, \cite{6921650}, \cite{10.1145/2872518.2889302}, \cite{varol2017online}, \cite{10.1016/j.ins.2016.08.036}, \cite{8093483}, \cite{8023129}, \cite{7918016}, \cite{10.1145/3110025.3110091}, \cite{8004887} and \cite{7837909} are included. Above mentioned researches have adopted a supervised learning approach toward this problem. Compared to the other categories, researches in this one have achieved highers scores. 
 
\section{The Data set}
There is no data set of bots and genuine accounts except one which was introduced in \cite{8946437} which does not include any Persian user account. As a result, a new data set suitable for present research's goals should have been gathered.

\subsection{Positive class}
Since there was no sample for the Instagram bot, a simplifying assumption was made: \textit{Since all harmful bots possess fake identities, gathering a set of fake accounts, will result in having a set of Instagram accounts which could be considered a superset for Instagram bots}. The same assumption has been made before in \cite{10.1145/3041021.3055135} when preparing a data set of twitter bots that were used for developing the famous bot detection tool Botometer in \cite{varol2017online} afterwords. Furthermore, to satisfy generalization considerations, five fake account providers for Instagram were referred to and 1000 fake accounts where ordered. 

\subsection{Negative class}
To have a set of genuine Instagram accounts, data from \textbf{Social Networks laboratory}\footnote{Faculty of Electrical and Computer Engineering - University of Tehran} is used. A data set of nearly all Persian Instagram accounts is gathered in this laboratory concerning the user's privacy. A random sample of size 1000 from this data was taken to represent genuine accounts. Since most user accounts in each social media belong to real humans, thus a random sample will provide a set of accounts including a large portion of genuine accounts.

In this data set, six features are used which are described in table \ref{table 1}.

\begin{table}[!ht]
\caption{Features introduced in the data set.}\smallskip
\centering
\smallskip\begin{tabular}{l|l}
\textbf{No} & \textbf{Feature name} \\
1 & username length\\
2 & full name length\\
3 & biography length\\
4 & followers count\\
5 & followings count\\
6 & Posts creation times \\
\end{tabular}
\label{table 1}
\end{table}

\section{Behavioral Modeling}
\noindent Behavior on Instagram has three forms: 1) to share a post, 2) to comment beneath a post and 3) to like a post or a comment; the three forms are referred to as behavioral properties. According to Instagram's structure, tracks of only the first property are accessible when checking a user's page and for having the rest, all data of the network is needed which is not ready here. As a result, only data describing users' posting behavior could be considered which corresponds to 6\textsuperscript{th} row in table \ref{table 1}. Since the number of posts differs among users, eight statistical measures are used to create an equal number of features for all users. Table \ref{table 2} illustrates names of above-mentioned measures.

\begin{table}[!ht]
\caption{Statistical measure used for modeling posts sharing times.}\smallskip
\centering
\smallskip\begin{tabular}{l|l}
\textbf{No} & \textbf{Statistical Measure} \\
1 & Min\\
2 & Max\\
3 & Mean\\
4 & Median\\
5 & Std\\
6 & Skewness \\
7 & Kurtosis \\
8 & Entropy \\
\end{tabular}
\label{table 2}
\end{table}

\subsection{ Effectiveness of Behavioral Modeling }
\noindent To test how effective this form of modeling is, a simple classifier is used, which is Gaussian Naive Bayes in this case, as a baseline. During two different experiments, one all features and then only basic features (all feature minus statistical measures defining behavioral properties) are fed to the classifier. In each experiment, the classifier is evaluated using 10-fold cross-validation. 
To check for the statistical significance of the reported results, the above-mentioned experiments are performed 1000 times which results in having 1,000 values for each of the measures per model. In each iteration (every single experiment of all 1,000), observations are shuffled 100 times to avoid bias. For 10-fold cross-validation, observations are shuffled again which sums up to 101 times random shuffling the data. It should be mentioned that the data is normalized using \textbf{Z-standardization} before being fed to the model.
The final results for this evaluation are represented in table \ref{table 3}. In this table, the mean of all values for each metric are reported plus or minus two times the standard deviation.

\begin{table}[!ht]
\caption{Behavioral modeling evaluation results.}\smallskip
\centering
\smallskip\begin{tabular}{l|l|l|l|l}
\textbf{No} & \textbf{Measure} & \textbf{Basic Features} & \textbf{All features} & \textbf{p-value} \\
\textbf{1} & \textbf{Accuracy} & 0.63 +/- 0.03 & 0.81 +/- 0.01 & 0.00\\
\textbf{2} & \textbf{Precision} & 0.61 +/- 0.05 & 0.99 +/- 0.00 & 0.00\\
\textbf{3} & \textbf{Recall} & 0.90 +/- 0.08 & 0.65 +/- 0.02 & 1.00\\
\textbf{4} & \textbf{F-1} & 0.71 +/- 0.03 & 0.78 +/- 0.01 & 0.00\\

\end{tabular}
\label{table 3}
\end{table}

\begin{figure}[!ht]
\centering
\includegraphics[width=0.4\textwidth]{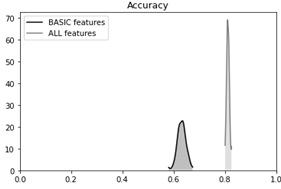} 
\caption{Distribution of \textbf{accuracy} values for both models. The darker curve (left) represents the model trained on all features and the lighter (right) has only used basic features.}
\label{fig1}
\end{figure}

\begin{figure}[!ht]
\centering
\includegraphics[width=0.4\textwidth]{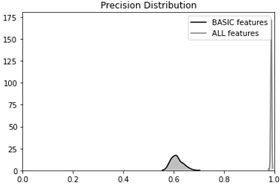} 
\caption{Distribution of \textbf{precision} values for both models. The darker curve (left) represents the model trained on all features and the lighter (right) has only used basic features.}
\label{fig2}
\end{figure}

\begin{figure}[!ht]
\centering
\includegraphics[width=0.4\textwidth]{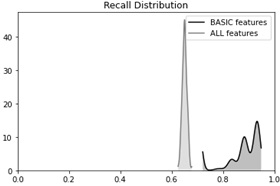} 
\caption{Distribution of \textbf{recall} values for both models. The darker curve (right) represents the model trained on all features and the lighter (left) has only used basic features.}
\label{fig3}
\end{figure}

\begin{figure}[!ht]
\centering
\includegraphics[width=0.4\textwidth]{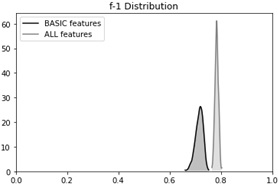} 
\caption{Distribution of \textbf{F-1} values for both models. The darker curve (left) represents the model trained on all features and the lighter (right) has only used basic features.}
\label{fig4}
\end{figure}

According to table \ref{table 3}, all measures show improvements but the recall. Although the recall is decreased when using all features, the increase in F-1 means that the precision is increased enough to support the decrease in the recall. Figures \ref{fig1} to \ref{fig4} illustrate distributions for metric values. 

Furthermore, the last column, depicts the result of a statistical test evaluating the null hypothesis which says \textit{the measure for both states are equal} against the alternative hypothesis saying \textit{numbers in the right column are greater than the left column} in a one-tailed test. As is illustrated in the right-most column, for all measures except the recall, the null hypothesis is strongly rejected in favor of the alternative hypothesis. To make certain that the experiment meets central limit theorem's criteria:

\begin{itemize}
    \item Observations for the test are picked randomly.
    \item Sample size is equal to 40, as a result, it is:
    \begin{itemize}
        \item less that 10\% of the population.
        \item greater that 30 (in order to support skewness in the data).
    \end{itemize}
\end{itemize}

From this experiment, it is clearly understood that behavioral modeling is effective in distinguishing bots and genuine accounts on Instagram, and this result is statistically significant.

\section{ Training the candidate models }

\subsection{ Features management }
\noindent In all machine learning studies, the quality of features must be examined before feeding the data to the model. In the present research, this job is done in two phases: 1) Calculating the importance of each feature 2) calculating the correlation between each pair of features.

After the two above-mentioned phases, redundant and highly correlated features (except one of them), also known as redundant features, are removed in order to reduce model complexity and training time. The remaining features are introduced in the table \ref{table 4}.

\begin{table}[!ht]
\caption{Remaining features after removing non-related and redundant features}\smallskip
\centering
\smallskip\begin{tabular}{l|l}
\textbf{No} & \textbf{Feature name} \\
1 & Max\\
2 & Std\\
3 & Skewness \\
4 & Entropy \\
5 & following count \\
6 & full name length \\
7 & biography length \\
\end{tabular}
\label{table 4}
\end{table}

\subsection{Finding candidate models}
\noindent Deep learning models have achieved astonishing results in the AI community, however, studies such as \cite{10.1145/3298689.3347058}, have shown that in many cases, fine-tuned classic machine learning algorithms (such as KNN, etc.) can achieve good performance even compared to deep learning methods. Furthermore, due to the complex structure of deep methods, a large amount of training data is required for them to achieve a high performance, which is not present in this study. Accordingly, classic machine learning methods are preferred over deep learning methods in current research.

Among classic methods, some are selected for further analysis including \textbf{KNN, Decision Tree, SVM, Random Forest, and AdaBoost}. This set, includes a spectrum of classic machine learning methods, from simple to complex. It is assumed that they can achieve good performance when using optimal hyperparameters.

\subsection{Fine-tuning the models}
\noindent In order to fine-tune the models, a portion of the data should be taken as the test set. A random sample including 30\% of all observations is selected for this process. A different sample is used for fine-tuning each model. In this research, sci-kit learn is used for implementing the solution, as a result, hyper-parameters names' are compatible with sci-kit learn the terminology. Final results in this phase are reported through tables \ref{table 5} to \ref{table 9}. 

\begin{table}[!ht]
\caption{Optimal hyper-parameter values for KNN}\smallskip
\centering
\smallskip\begin{tabular}{l|l}
\textbf{Name} & \textbf{Value} \\
Algorithm & Ball Tree\\
Leave size & 10\\
K & 5 \\
Distance & Manhattan \\
Weights & Distance \\
\end{tabular}
\label{table 5}
\end{table}

\begin{table}[!ht]
\caption{Optimal hyper-parameter values for Decision Tree}\smallskip
\centering
\smallskip\begin{tabular}{l|l}
\textbf{Name} & \textbf{Value} \\
Criterion & Gini-index\\
Max depth & 10\\
Min samples leaf & 2 \\
Min impurity split & 3 \\
Splitter & Random \\
\end{tabular}
\label{table 6}
\end{table}

\begin{table}[!ht]
\caption{Optimal hyper-parameter values for SVM}\smallskip
\centering
\smallskip\begin{tabular}{l|l}
\textbf{Name} & \textbf{Value} \\
C & 3\\
Kernel & Polynomial\\
degree & 5 \\
coef0 & 1.5 \\
shrinking & True \\
probabilistic approximations & True \\
\end{tabular}
\label{table 7}
\end{table}

\begin{table}[!ht]
\caption{Optimal hyper-parameter values for Random Forest}\smallskip
\centering
\smallskip\begin{tabular}{l|l}
\textbf{Name} & \textbf{Value} \\
Criterion & Entropy\\
Max depth & None\\
Max features & 5 \\
Min samples split & 3 \\
n estimators & 20 \\
\end{tabular}
\label{table 8}
\end{table}

\begin{table}[!ht]
\caption{Optimal hyper-parameter values for AdaBoost}\smallskip
\centering
\smallskip\begin{tabular}{l|l}
\textbf{Name} & \textbf{Value} \\
Algorithm & SAMME\\
Learning rate & 1.0\\
n estimators & 50 \\
\end{tabular}
\label{table 9}
\end{table}

\subsection{Train and evaluate}
\noindent When optimal hyper-parameters are determined, fine-tuned models should be trained with enough data. For each model, the whole data is shuffled, a random sample including 70\% of observations is selected for training models and the rest is preserved for the test. In addition to test data, the models are evaluated using 10-fold cross-validation, however, reported values for all metrics are from cross-validation except for the area under the ROC curve which is computed using the test data. In this table, the greatest values are written in bold. The evaluation results are illustrated in \ref{table10}.

\begin{table*}[!ht]
\centering
\caption{Classification Results}\smallskip
\begin{tabular}{l|l|l|l|l|l}
 \textbf{The model} & \textbf{Accuracy} & \textbf{Precision} & \textbf{Recall} & \textbf{F-1} &
 \textbf{ROC AUC} \\
KNN & 0.92 & 0.95 & 0.90 & 0.92 & 0.97 \\
Decision Tree & 0.92 & 0.93 & 0.91 & 0.92 & 0.94 \\
SVM & 0.93 & 0.94 & 0.94 & 0.94 & 0.96 \\
Random Forest & \textbf{0.95} & 0.94 & 0.94 & \textbf{0.95} & \textbf{0.99} \\
AdaBoost & \textbf{0.95} & \textbf{0.95} & \textbf{0.95} & \textbf{0.95} & \textbf{0.99} \\
\end{tabular}
\label{table10}
\end{table*}

According to the results, the \textbf{AdaBoost} has achieved the highest scores. Compared to the baseline model, this model has achieved a better score in metrics such as accuracy, recall, and f-1 but has got a lower score in precision. 

\section{AdaBoost further analysis}
\subsection{Classification confidence}
\noindent To examine how confident the AdaBoost is when it makes decisions about observations, \textbf{probabilities of being a bot} computed by the model for the test data are computed. 

\begin{figure}[!ht]
\centering
\includegraphics[width=0.4\textwidth]{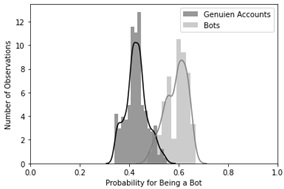} 
\caption{Distributions of probabilities for being a bot computed for the test data.}
\label{fig5}
\end{figure}

In this figure, the curve on the left (the darker) depicts the distribution of probabilities for being a bot computed for genuine accounts in the test data. The curve on the right (the lighter one) shows the same probabilities for genuine accounts in the test data. Ideally, both curves should have been located away from each other, however, the figure shows the opposite. Consequently, it is understood that despite its high accuracy, the model is not much confident about the decisions it makes. 

To quantify the results of this experiment, 95\% confidence intervals are computed for both distributions. To do this, one sample should be picked from both bots and genuine accounts in the test data. Central limit theorem's criteria should be satisfied when using confidence intervals, and to do so, the following conditions must hold for the selected samples.

\begin{itemize}
    \item The observations must have been selected uniformly at random.
    \item Sample's size must be larger than 30 (this number is most likely to be enough since neither of the distributions is highly skewed).
    \item Sample's size must be less than 10\% of the population.
\end{itemize}

Information describing the two intervals are provided in table \ref{table 11}.

\begin{table}[!ht]
\centering
\caption{Confidence interval information}\smallskip
\begin{tabular}{l|l}
 \textbf{Class} & \textbf{Confidence interval} \\
Bots & 0.58 +/- 0.016 \\
Genuine Accounts & 0.43 +/- 0.014 \\
\end{tabular}
\label{table 11}
\end{table}

Although the confidence intervals reported in table \ref{table 11} support the idea that the two distributions are close, the thin marginal errors in both intervals show that they do not overlap. This experiment, weakens the idea of overlapping distributions thus supporting effectiveness of the AdaBoost in this problem. This phenomenon is believed to be existed due to the small number of features used for describing the data.

\subsection{Features' importance and relations}
\noindent One way to analyze decisions made by a model is to compute the importance of each feature in the classification process (recall some features were removed in the feature management phase, thus Gini importance is computed for remaining features by the model). One method for computing the importance values, is \textbf{gini importance} which is described in \cite{10.5555/2999611.2999660}. The Gini importance values for all features which are used in the data are computed and reported in figure 6.

\begin{figure*}[!ht]
\centering
\includegraphics[width=0.8\textwidth]{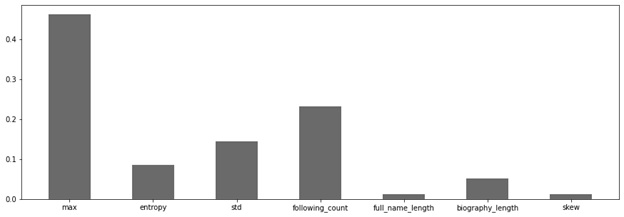} 
\caption{Gini importance for all features which are used in classification process.}
\label{fig6}
\end{figure*}

According to this figure, \textbf{Max} is the most effective feature in the classification process. It is followed by \textbf{following count}, \textbf{standard deviation}, \textbf{entropy} and three others. Among the four most important features, three are behavioral features. This is another proof of the effectiveness of behavioral features.

\begin{figure}[!ht]
\centering
\includegraphics[width=0.4\textwidth]{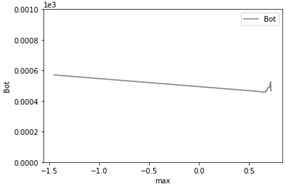} 
\caption{The relation between values of \textbf{Max} feature and the probability of being a bot.}
\label{fig7}
\end{figure}

\begin{figure}[!ht]
\centering
\includegraphics[width=0.4\textwidth]{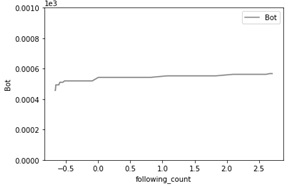} 
\caption{The relation between values of \textbf{Following Count} feature and the probability of being a bot.}
\label{fig8}
\end{figure}

\begin{figure}[!ht]
\centering
\includegraphics[width=0.4\textwidth]{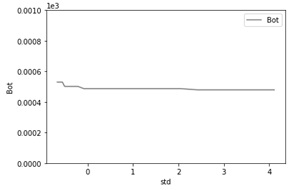} 
\caption{The relation between values of \textbf{std} feature and the probability of being a bot.}
\label{fig9}
\end{figure}

\begin{figure}[!ht]
\centering
\includegraphics[width=0.4\textwidth]{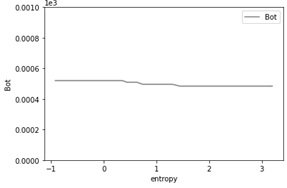} 
\caption{The relation between values of \textbf{Entropy} feature and the probability of being a bot.}
\label{fig10}
\end{figure}

The last step toward investigating the AdaBoost is to analyze relations between every single feature's values and the probability of being a bot which is computed by the model. This is done by using \textbf{Partial Dependency Plots (PDPs)}; these plots are illustrated in figures \ref{fig7} to \ref{fig10}. Since the rest of the features have rather no observable relation with the probability of being a bot, their corresponding PDPs are not presented. 

\subsection{Final experiment - I}
\noindent A final experiment is designed to investigate the practical effectiveness of the AdaBoost. This experiment includes the following steps:
\begin{itemize}
    \item A random sample of size 5000 is selected from social networks lab's data
    \item All observations from the sample are fed to the AdaBoost and the probability of being a bot is computed for all of them.
    \item Observations are sorted in decreasing order of probabilities and 50 first accounts were selected for hand-check.
\end{itemize}

The results of this experiment are briefly illustrated in table \ref{table 12} and are comprehensively described here. Among these fifty accounts: 
\begin{itemize}
    \item Ten were not found. This might have been due to username change and/or account removal. 
    \item Ten accounts were similar to genuine accounts. This experiment is based on observing profile pictures, the similarity between the profile picture and other posted pictures, etc. It is clear that in this step, decisions are made based on the user's content. Since content-related features are absent in the data set, such a phenomenon is likely to be observed.
    \item Thirty accounts looked like genuine accounts. Decisions in the step, are made on a similar basis as the previous step. Recalling that content-related features are absent, this observation shows that behavioral features are in tight relation with content features.
\end{itemize}

\begin{table}[!ht]
\centering
\caption{Result of the hand-check for random accounts}\smallskip
\begin{tabular}{l|l}
 \textbf{Count} & \textbf{Status} \\
10 & Account was not found. \\
10 & Accounts which look like genuine accounts. \\
30 & Potential Bots.\\
\end{tabular}
\label{table 12}
\end{table}

\subsection{Final experiment - II}
\noindent One way to check the effectiveness of classification is to evaluate the classifier for distinguishing different types of accounts. To investigate the classifier in this fashion, an experiment was designed. This experiment includes five iterations. In each iteration, accounts belonging to one of the visited Instagram service providers are hold out plus an equal number of genuine accounts altogether as test data, and the model (AdaBoost with optimum hyper-parameters) is trained on rest of data. The results of this experiment are reported in table \ref{table 13}.

\begin{table}[!ht]
\centering
\caption{Results of the class by class evaluation}\smallskip
\begin{tabular}{l|l|l|l|l|l}
 \textbf{Measures} & \textbf{Itr 1} & \textbf{Itr 2} & \textbf{Itr 3} & \textbf{Itr 4} & \textbf{Itr 5} \\
\textbf{Accuracy} & 0.98 & 0.98 & 0.96 & 0.95 & 0.97 \\
\textbf{Precision} & 0.98 & 0.98 & 0.96 & 0.95 & 0.97 \\
\textbf{Recall} & 0.98 & 0.98 & 0.96 & 0.95 & 0.97\\
\textbf{F-1} & 0.98 & 0.98 & 0.96 & 0.95 & 0.97
\end{tabular}
\label{table 13}
\end{table}

In table \ref{table 13}, each columns includes same values for all metrics. Although this observation seems strange, it is similar to results in table \ref{table10} thus the results are reasonable in this setting.

Furthermore, in all iterations, except for iteration 4, all metrics' values are greater than 0.95 which was achieved in 10-fold cross-validation (as reported in table \ref{table10}). Such an increase in evaluation metrics might be observed due to the following reasons:
\begin{itemize}
    \item Usually, evaluation metrics achieve higher scores with test data.
    \item In this experiment, data is re-sampled with replacement to provide an equal size for both classes. Extra observation gathered in this way, might have improved classification's quality.
\end{itemize}

\section{Results}
\noindent In this research, a new data set including samples of bots and genuine Instagram accounts are gathered concerning generality concerns. Behavioral modeling is introduced as an effective technique for modeling Instagram users' data concerned with the bot detection problem. Classic machine learning algorithms are used for classification and are fine-tuned to have their performance boosted. It is shown that fine-tuned classic machine learning models perform well in the current study. Boosting, as an ensemble method has proved to be more effective compared to simpler methods. AdaBoost, which has achieved the best performance among all studied methods. Creation times of most recent posts (which is introduced as the \textbf{Max} in the data set) are proved to be the most important feature in classification which has a negative linear relationship with the probability of being a bot computed by the classifier. 

\section{ Acknowledgments}
This work would have been much more difficult without helps from others. People who have had significant roles in this research are mentioned bellow:

\noindent\textbf{Dr Behnam Bahrak}: Counsellor on statistical issues.\\
\textbf{Abbas Ma'allahi}: Developing the crawling system for Instagram.\\
\textbf{Mitham Saki}:Developing and maintaining the data cluster at Social Networks Lab.\\
\textbf{Seyed Morteza Ghavami}:Developing processes which maintain the data in the cluste.\\
\textbf{Milad Sayadamooz}: Developing web services for working with the data storage cluster.\\

\bibliography{refs.bib}

\begin{thebibliography}{}

\bibitem[\protect\citeauthoryear{{Akyon} and {Esat Kalfaoglu}}{2019}]{8946437}
{Akyon}, F.~C., and {Esat Kalfaoglu}, M.
\newblock 2019.
\newblock Instagram fake and automated account detection.
\newblock In {\em 2019 Innovations in Intelligent Systems and Applications
  Conference (ASYU)},  1--7.

\bibitem[\protect\citeauthoryear{Alarifi, Alsaleh, and
  Al-Salman}{2016}]{10.1016/j.ins.2016.08.036}
Alarifi, A.; Alsaleh, M.; and Al-Salman, A.
\newblock 2016.
\newblock Twitter turing test.
\newblock {\em Inf. Sci.} 372(C):332–346.

\bibitem[\protect\citeauthoryear{{Cai}, {Li}, and {Zengi}}{2017}]{8004887}
{Cai}, C.; {Li}, L.; and {Zengi}, D.
\newblock 2017.
\newblock Behavior enhanced deep bot detection in social media.
\newblock In {\em 2017 IEEE International Conference on Intelligence and
  Security Informatics (ISI)},  128--130.

\bibitem[\protect\citeauthoryear{Carminati, Ferrari, and
  Viviani}{2014}]{Carminati2013}
Carminati, B.; Ferrari, E.; and Viviani, M.
\newblock 2014.
\newblock {\em {Security and Trust in Online Social Networks}}.
\newblock Morgan \& Claypool Publishers.

\bibitem[\protect\citeauthoryear{{Chavoshi}, {Hamooni}, and
  {Mueen}}{2016}]{7837909}
{Chavoshi}, N.; {Hamooni}, H.; and {Mueen}, A.
\newblock 2016.
\newblock Debot: Twitter bot detection via warped correlation.
\newblock In {\em 2016 IEEE 16th International Conference on Data Mining
  (ICDM)},  817--822.

\bibitem[\protect\citeauthoryear{{Chu} \bgroup et al\mbox.\egroup
  }{2012}]{6280553}
{Chu}, Z.; {Gianvecchio}, S.; {Wang}, H.; and {Jajodia}, S.
\newblock 2012.
\newblock Detecting automation of twitter accounts: Are you a human, bot, or
  cyborg?
\newblock {\em IEEE Transactions on Dependable and Secure Computing}
  9(6):811--824.

\bibitem[\protect\citeauthoryear{Cresci \bgroup et al\mbox.\egroup
  }{2017}]{10.1145/3041021.3055135}
Cresci, S.; Di~Pietro, R.; Petrocchi, M.; Spognardi, A.; and Tesconi, M.
\newblock 2017.
\newblock The paradigm-shift of social spambots: Evidence, theories, and tools
  for the arms race.
\newblock In {\em Proceedings of the 26th International Conference on World
  Wide Web Companion}, WWW ’17 Companion,  963–972.
\newblock Republic and Canton of Geneva, CHE: International World Wide Web
  Conferences Steering Committee.

\bibitem[\protect\citeauthoryear{{Cross} and {Jain}}{1983}]{4767341}
{Cross}, G.~R., and {Jain}, A.~K.
\newblock 1983.
\newblock Markov random field texture models.
\newblock {\em IEEE Transactions on Pattern Analysis and Machine Intelligence}
  PAMI-5(1):25--39.

\bibitem[\protect\citeauthoryear{Dacrema, Cremonesi, and
  Jannach}{2019}]{10.1145/3298689.3347058}
Dacrema, M.~F.; Cremonesi, P.; and Jannach, D.
\newblock 2019.
\newblock Are we really making much progress? a worrying analysis of recent
  neural recommendation approaches.
\newblock In {\em Proceedings of the 13th ACM Conference on Recommender
  Systems}, RecSys ’19,  101–109.
\newblock New York, NY, USA: Association for Computing Machinery.

\bibitem[\protect\citeauthoryear{Danezis and
  Mittal}{2009}]{Danezis_sybilinfer:detecting}
Danezis, G., and Mittal, P.
\newblock 2009.
\newblock Sybilinfer: Detecting sybil nodes using social networks.

\bibitem[\protect\citeauthoryear{Davis \bgroup et al\mbox.\egroup
  }{2016}]{10.1145/2872518.2889302}
Davis, C.~A.; Varol, O.; Ferrara, E.; Flammini, A.; and Menczer, F.
\newblock 2016.
\newblock Botornot: A system to evaluate social bots.
\newblock In {\em Proceedings of the 25th International Conference Companion on
  World Wide Web}, WWW ’16 Companion,  273–274.
\newblock Republic and Canton of Geneva, CHE: International World Wide Web
  Conferences Steering Committee.

\bibitem[\protect\citeauthoryear{{Dickerson}, {Kagan}, and
  {Subrahmanian}}{2014}]{6921650}
{Dickerson}, J.~P.; {Kagan}, V.; and {Subrahmanian}, V.~S.
\newblock 2014.
\newblock Using sentiment to detect bots on twitter: Are humans more
  opinionated than bots?
\newblock In {\em 2014 IEEE/ACM International Conference on Advances in Social
  Networks Analysis and Mining (ASONAM 2014)},  620--627.

\bibitem[\protect\citeauthoryear{Gao \bgroup et al\mbox.\egroup
  }{2015}]{Gao2015SybilFrameAD}
Gao, P.; Gong, N.~Z.; Kulkarni, S.~R.; Thomas, K.; and Mittal, P.
\newblock 2015.
\newblock Sybilframe: A defense-in-depth framework for structure-based sybil
  detection.
\newblock {\em ArXiv} abs/1503.02985.

\bibitem[\protect\citeauthoryear{Gilani, Kochmar, and
  Crowcroft}{2017}]{10.1145/3110025.3110091}
Gilani, Z.; Kochmar, E.; and Crowcroft, J.
\newblock 2017.
\newblock Classification of twitter accounts into automated agents and human
  users.
\newblock In {\em Proceedings of the 2017 IEEE/ACM International Conference on
  Advances in Social Networks Analysis and Mining 2017}, ASONAM ’17,
  489–496.
\newblock New York, NY, USA: Association for Computing Machinery.

\bibitem[\protect\citeauthoryear{Gong, Frank, and
  Mittal}{2014}]{DBLP:journals/tifs/GongFM14}
Gong, N.~Z.; Frank, M.; and Mittal, P.
\newblock 2014.
\newblock Sybilbelief: {A} semi-supervised learning approach for
  structure-based sybil detection.
\newblock {\em {IEEE} Trans. Information Forensics and Security} 9(6):976--987.

\bibitem[\protect\citeauthoryear{{Jia}, {Wang}, and {Gong}}{2017}]{8023129}
{Jia}, J.; {Wang}, B.; and {Gong}, N.~Z.
\newblock 2017.
\newblock Random walk based fake account detection in online social networks.
\newblock In {\em 2017 47th Annual IEEE/IFIP International Conference on
  Dependable Systems and Networks (DSN)},  273--284.

\bibitem[\protect\citeauthoryear{Jiang \bgroup et al\mbox.\egroup
  }{2013}]{10.1145/2517040}
Jiang, J.; Wilson, C.; Wang, X.; Sha, W.; Huang, P.; Dai, Y.; and Zhao, B.~Y.
\newblock 2013.
\newblock Understanding latent interactions in online social networks.
\newblock {\em ACM Trans. Web} 7(4).

\bibitem[\protect\citeauthoryear{{Kantepe} and {Ganiz}}{2017}]{8093483}
{Kantepe}, M., and {Ganiz}, M.~C.
\newblock 2017.
\newblock Preprocessing framework for twitter bot detection.
\newblock In {\em 2017 International Conference on Computer Science and
  Engineering (UBMK)},  630--634.

\bibitem[\protect\citeauthoryear{Lee, Eoff, and
  Caverlee}{2011}]{Lee11sevenmonths}
Lee, K.; Eoff, B.~D.; and Caverlee, J.
\newblock 2011.
\newblock Seven months with the devils: a long-term study of content polluters
  on twitter.
\newblock In {\em In AAAI Int’l Conference on Weblogs and Social Media
  (ICWSM}.

\bibitem[\protect\citeauthoryear{Leskovec \bgroup et al\mbox.\egroup
  }{2008}]{Leskovec08communitystructure}
Leskovec, J.; Lang, K.~J.; Dasgupta, A.; and Mahoney, M.~W.
\newblock 2008.
\newblock Community structure in large networks: Natural cluster sizes and the
  absence of large well-defined clusters.

\bibitem[\protect\citeauthoryear{Louppe \bgroup et al\mbox.\egroup
  }{2013}]{10.5555/2999611.2999660}
Louppe, G.; Wehenkel, L.; Sutera, A.; and Geurts, P.
\newblock 2013.
\newblock Understanding variable importances in forests of randomized trees.
\newblock In {\em Proceedings of the 26th International Conference on Neural
  Information Processing Systems - Volume 1}, NIPS’13,  431–439.
\newblock Red Hook, NY, USA: Curran Associates Inc.

\bibitem[\protect\citeauthoryear{{Mehrotra}, {Sarreddy}, and
  {Singh}}{2016}]{7918016}
{Mehrotra}, A.; {Sarreddy}, M.; and {Singh}, S.
\newblock 2016.
\newblock Detection of fake twitter followers using graph centrality measures.
\newblock In {\em 2016 2nd International Conference on Contemporary Computing
  and Informatics (IC3I)},  499--504.

\bibitem[\protect\citeauthoryear{Mohaisen, Yun, and
  Kim}{2010}]{10.1145/1879141.1879191}
Mohaisen, A.; Yun, A.; and Kim, Y.
\newblock 2010.
\newblock Measuring the mixing time of social graphs.
\newblock In {\em Proceedings of the 10th ACM SIGCOMM Conference on Internet
  Measurement}, IMC ’10,  383–389.
\newblock New York, NY, USA: Association for Computing Machinery.

\bibitem[\protect\citeauthoryear{Murphy, Weiss, and
  Jordan}{2013}]{DBLP:journals/corr/abs-1301-6725}
Murphy, K.~P.; Weiss, Y.; and Jordan, M.~I.
\newblock 2013.
\newblock Loopy belief propagation for approximate inference: An empirical
  study.
\newblock {\em CoRR} abs/1301.6725.

\bibitem[\protect\citeauthoryear{{Subrahmanian} \bgroup et al\mbox.\egroup
  }{2016}]{7490315}
{Subrahmanian}, V.~S.; {Azaria}, A.; {Durst}, S.; {Kagan}, V.; {Galstyan}, A.;
  {Lerman}, K.; {Zhu}, L.; {Ferrara}, E.; {Flammini}, A.; and {Menczer}, F.
\newblock 2016.
\newblock The darpa twitter bot challenge.
\newblock {\em Computer} 49(6):38--46.

\bibitem[\protect\citeauthoryear{Varol \bgroup et al\mbox.\egroup
  }{2017}]{varol2017online}
Varol, O.; Ferrara, E.; Davis, C.~A.; Menczer, F.; and Flammini, A.
\newblock 2017.
\newblock Online human-bot interactions: Detection, estimation, and
  characterization.
\newblock In {\em International AAAI Conference on Web and Social Media},
  280{\textendash}289.
\newblock AAAI.

\bibitem[\protect\citeauthoryear{Viswanath \bgroup et al\mbox.\egroup
  }{2010}]{10.1145/1851275.1851226}
Viswanath, B.; Post, A.; Gummadi, K.~P.; and Mislove, A.
\newblock 2010.
\newblock An analysis of social network-based sybil defenses.
\newblock {\em SIGCOMM Comput. Commun. Rev.} 40(4):363–374.

\bibitem[\protect\citeauthoryear{Wang \bgroup et al\mbox.\egroup
  }{2013}]{Wang2013SocialTT}
Wang, G.; Mohanlal, M.; Wilson, C.; Wang, X.; Metzger, M.~J.; Zheng, H.; and
  Zhao, B.~Y.
\newblock 2013.
\newblock Social turing tests: Crowdsourcing sybil detection.
\newblock {\em ArXiv} abs/1205.3856.

\bibitem[\protect\citeauthoryear{Yang \bgroup et al\mbox.\egroup
  }{2014}]{10.1145/2556609}
Yang, Z.; Wilson, C.; Wang, X.; Gao, T.; Zhao, B.~Y.; and Dai, Y.
\newblock 2014.
\newblock Uncovering social network sybils in the wild.
\newblock {\em ACM Trans. Knowl. Discov. Data} 8(1).

\bibitem[\protect\citeauthoryear{{Yu} \bgroup et al\mbox.\egroup
  }{2008a}]{4531141}
{Yu}, H.; {Gibbons}, P.~B.; {Kaminsky}, M.; and {Xiao}, F.
\newblock 2008a.
\newblock Sybillimit: A near-optimal social network defense against sybil
  attacks.
\newblock In {\em 2008 IEEE Symposium on Security and Privacy (sp 2008)},
  3--17.

\bibitem[\protect\citeauthoryear{{Yu} \bgroup et al\mbox.\egroup
  }{2008b}]{4542826}
{Yu}, H.; {Kaminsky}, M.; {Gibbons}, P.~B.; and {Flaxman}, A.~D.
\newblock 2008b.
\newblock Sybilguard: Defending against sybil attacks via social networks.
\newblock {\em IEEE/ACM Transactions on Networking} 16(3):576--589.

\bibitem[\protect\citeauthoryear{Zhang \bgroup et al\mbox.\egroup
  }{2016}]{6fea7744517d4694ad9dc535b9f37dc7}
Zhang, J.; Zhang, R.; Sun, J.; Zhang, Y.; and Zhang, C.
\newblock 2016.
\newblock Truetop: A sybil-resilient system for user influence measurement on
  twitter.
\newblock {\em IEEE/ACM Transactions on Networking} 24(5):2834--2846.

\end{thebibliography}
\bibliographystyle{aaai}
\end{document}